\documentclass[aps,prstab,twocolumn,showpacs,superscriptaddress,groupedaddress]{revtex4}  

\usepackage{dcolumn}   
\usepackage{bm}        
\usepackage{amssymb}   
\usepackage{color}

\newcommand{\secref}[1]{Section~\ref{#1}}
\newcommand{\figref}[1]{Fig.~\ref{#1}}
\newcommand{\tblref}[1]{Table~\ref{#1}}
\newcommand{\eqnref}[1]{(\ref{#1})}

\newcommand{\angstrom}{\mbox{\normalfont\AA}}
\newcommand{\exn}{\ensuremath{\epsilon_{xn}}}

\newcommand{\urad}{\ensuremath{\mu\textrm{rad}}}
\usepackage{color}

\usepackage{amsmath}
\usepackage{verbatim} 
\usepackage{color}
\usepackage{float}
\usepackage{graphicx}  
\usepackage{epstopdf}

\usepackage[caption=false,font=footnotesize]{subfig}

\begin{document}
\title{Nano-modulated electron beams via electron diffraction and emittance exchange for coherent x-ray generation}
\author{E.A.~Nanni}\affiliation{SLAC National Accelerator Laboratory, Menlo Park, CA 94025, USA} \affiliation{Massachusetts Institute of Technology, Cambridge, MA 02139, USA}
\author{W.S.~Graves}\affiliation{Arizona State University, Tempe, Arizona 85287, USA}
\author{D.E.~Moncton} \affiliation{Massachusetts Institute of Technology, Cambridge, MA 02139, USA}

%
%
 \noaffiliation%
\vskip 0.25cm

\date{\today}

\begin{abstract}
We present a new method for generation of relativistic electron beams with current modulation on the nanometer scale and below.  The current modulation is produced by diffracting relativistic electrons in single crystal Si, accelerating the diffracted beam and imaging the crystal structure, then transferring the image into the temporal dimension via emittance exchange. The modulation period can be tuned by adjusting electron optics after diffraction. This tunable longitudinal modulation can have a period as short as a few angstroms, enabling production of coherent hard x-rays from a source based on inverse Compton scattering with total accelerator length of approximately ten meters. Electron beam simulations from cathode emission through diffraction, acceleration and image formation with variable magnification are presented along with estimates of the coherent x-ray output properties.

\end{abstract}

\maketitle

\section{Introduction}

Hard x-ray free-electron lasers (FELs) such as LCLS, SACLA, and XFEL \cite{emma2010first,pile2011x,altarelli2006european} rely on long linear accelerators to produce high energy ($>7$ GeV) electron beams that meet the resonant condition for the x-ray wavelength $ \lambda_x \sim \lambda_u / \gamma^2$ for output from an undulator with period $\lambda_u$ of a few centimeters and kinetic energy $E_k \approx \gamma mc^2$.   The high electron energy required has the advantages of lowered electron beam geometric emittance becoming less than the diffraction-limited mode area of the coherent x-ray beam, and reduced space charge forces that might interfere with FEL gain.  The drawbacks are that the required facilities are large and expensive with just a few contemplated around the world, and that the electron modulation and the resulting x-ray beams produced by self-amplified spontaneous emission (SASE) are not fully coherent. FEL facilities are investigating several schemes in order to achieve greater temporal coherence including self-seeding \cite{feldhaus1997possible,saldin2001x,ding2010two}, high-gain harmonic generation \cite{ben1991proposed,yu1991generation,yu2000high} and echo-enhanced harmonic generation \cite{xiang2009echo,zhao2012first}. With harmonic generation, a seed laser at a longer wavelength is used to initiate the modulation on the electron bunch which then radiates at a harmonic of the laser wavelength with results demonstrated at extreme UV and soft x-ray wavelengths. Self-seeding uses a small portion of the SASE spectrum to seed a coherent pulse in a subsequent undulator. These techniques continue to require the use of GeV electron beams, include the addition of undulators and dispersive sections; and have proved challenging to scale to hard x-ray wavelengths. 

Transmission electron microscopes (TEMs) meanwhile produce a coherent spatial electron modulation at the scale of angstroms and do so using electrons with much lower energies of a few hundred keV from a compact device.  This demonstrates that an ultra-relativistic beam is not \textit{a priori} necessary for the angstrom-scale electron modulation typically produced by an x-ray FEL. Furthermore, this modulation can have a much greater coherence.  We take advantage of TEM-like electron diffraction of a modestly relativistic 7 MeV electron beam from a silicon target in combination with a previous concept \cite{graves2012intense,graves2013compact} for transforming a spatial modulation to a coherent temporal modulation at short wavelength to produce an electron beam suitable for coherent x-ray generation. Our earlier work relied on generation of an electron beam from a nanostructured cathode that was limited to scales of hundreds of nanometers or longer.  The current concept has several important advantages including potential for hard x-rays set by the atomic-scale limits of the diffracting crystal, mitigation of space charge effects due to diffraction at relativistic energy and use of a robust conventional flat cathode to produce the electrons. 

{The target that will be used to produce a transverse density modulation in the electron bunch is a silicon grating, with the grating pattern approximately normal to the electron beam's direction of propagation. As shown in \secref{sec:difcontrast}, the varying thickness of the silicon results in a spatially varying modulation of the probability for the electron to be diffracted by a particular Bragg peak  in the basic silicon structure. The thickness parameters of the grating can be chosen to optimize this diffraction contrast using the effect of dynamical extinction. After the target, either the Bragg-scattered electrons or forward-scattered electrons are accelerated and sent through the imaging optics. With this diffraction-contrast imaging, the limitation on the modulation period set at the grating is the minimum feature size that can be fabricated, of order 100~nm for standard lithographic techniques. However, this modulation period can be magnified or demagnified downstream with standard magnetic optics. Here we show that it is possible to demagnify the modulation period by a factor of 150 producing modulation periods of one nm. This approach can be extended to modulation at the Angstrom scale, as discussed in \secref{sec:phasecontrast}, with a uniform target thickness and the interference of multiple diffracted beams, similar to phase-contrast imaging used in high-voltage electron microscopy.}

\begin{figure*}[t]
  \centering
{\includegraphics[trim=0cm 0cm 0cm 0cm, clip=true, width=0.99\textwidth]{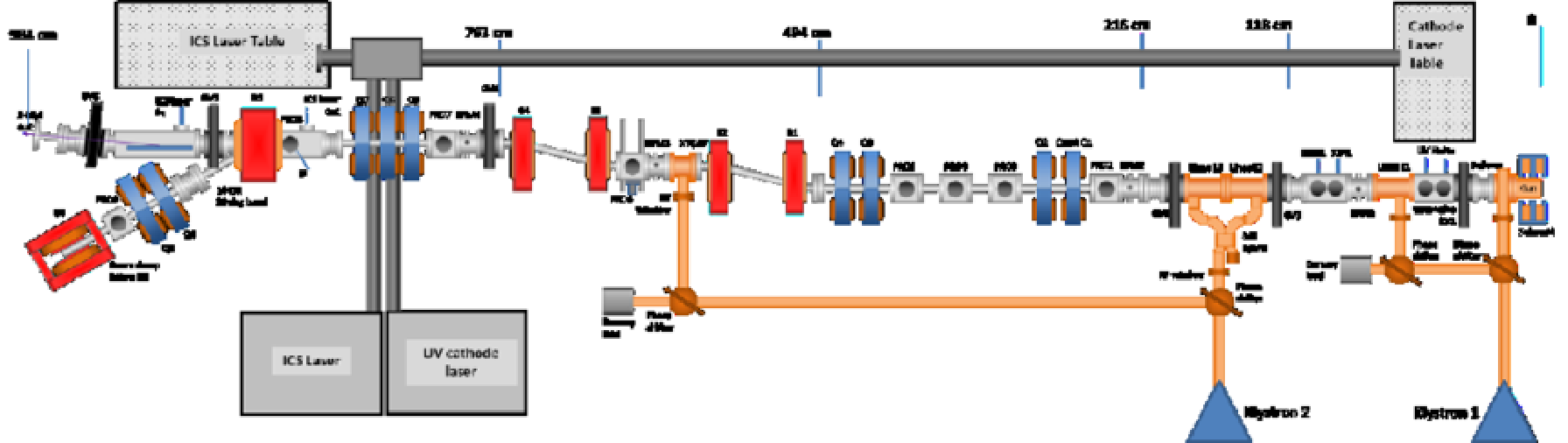} 
	\caption {Schematic of the compact coherent x-ray source with RF photo-injector, electron diffraction crystal, X-band linac, EEX line and ICS laser interaction area. Entire assembly is approximately 10~m long.}
  \label{fig:layout}}
\end{figure*}

A hard x-ray FEL based on this technology would fit comfortably in existing industrial, academic and medical laboratories at a cost comparable to other sophisticated analytical instruments. A schematic of the coherent x-ray source is shown in \figref{fig:layout} with a total length of approximately 10~m. We estimate that the output pulse for this source will have about 10~nJ of energy compared to the millijoule levels of the large machines.  However, it will be fully coherent, unlike SASE, and would be an excellent source for seeding temporally coherent x-rays from the large machines as well as directly producing new science opportunities for a broad range of applications.

\begin{figure*}[t]
  \centering
  \subfloat[]{\includegraphics[width=0.28\textwidth]{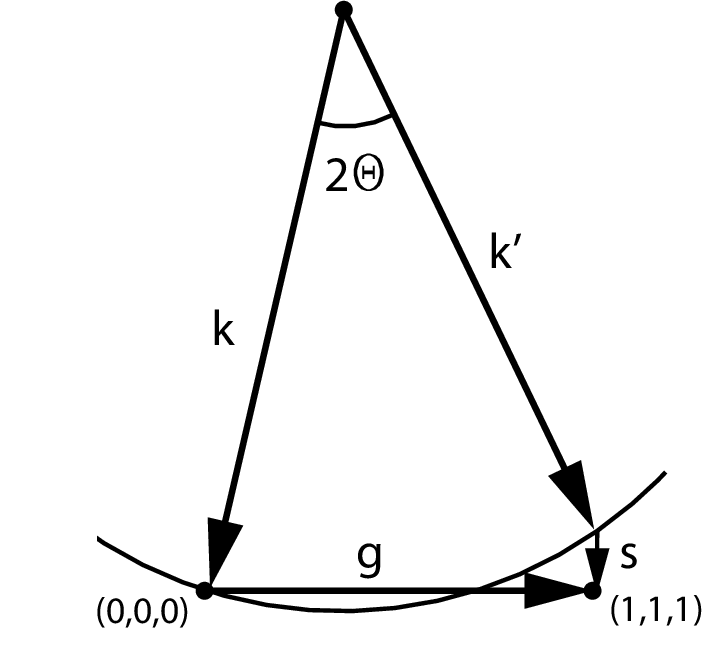}
  \label{fig:side}}
		\subfloat[]{\includegraphics[width=0.36\textwidth]{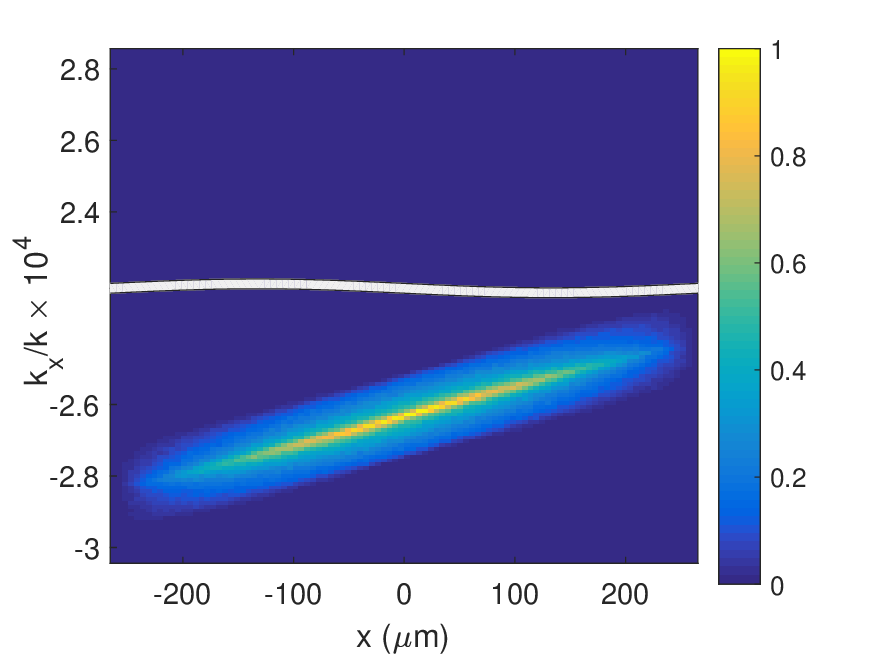}
	\label{fig:b0diffcontrast}}
	\subfloat[]{\includegraphics[width=0.36\textwidth]{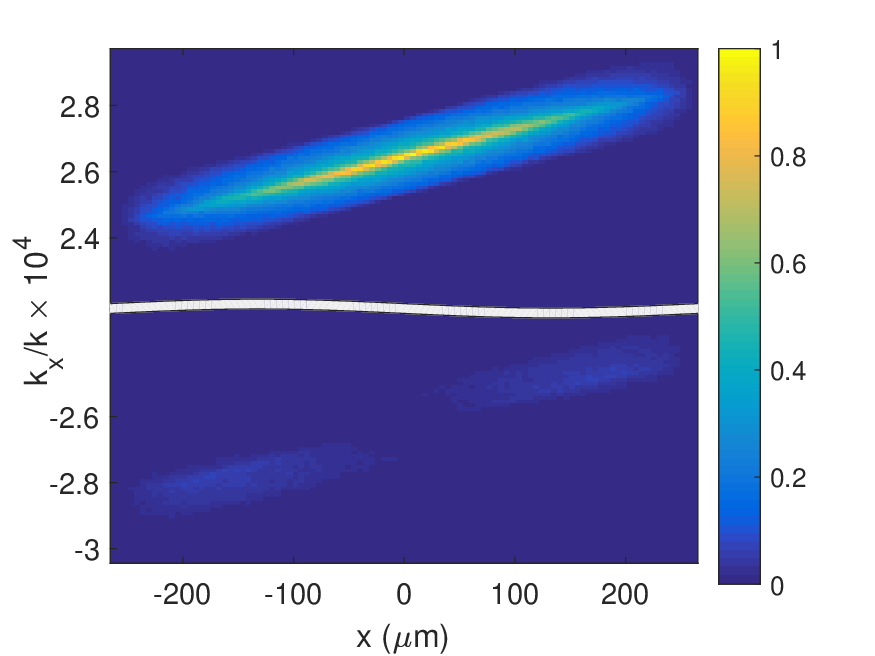}
	\label{fig:4closeup}}
	\caption {(a) Scattering geometry for an incident particle. (b) The phase space of the incident electron bunch immediately prior to the target.  (c) The calculated diffraction for the electron bunch from the RF photo-injector plotted as the relative density of electrons in the transverse phase space. With the thickness of the single crystal Si at $t={\xi }/2$, 98$\%$ of the beam  undergoes one scattering event (upper panel). The remaining electrons (lower panel) did not scatter. }
  \label{fig:kspace}
\end{figure*}

\section{Electron Diffraction}	

Ultra-fast electron diffraction experiments using electron bunches produced by an RF photo-injector with an electron energy on the order of a few MeV and a charge of several pC \cite{murooka2011transmission,hastings2006ultrafast,musumeci2008relativistic,hada2012regae,miller2014mapping} have demonstrated the feasibility of obtaining useful diffraction patterns with low emittance and high peak current electron beams. The main constraint for the use of an RF photo-injector for electron diffraction is the ability to achieve a low enough emittance or angular spread of the incoming electron bunch. Additionally, the diffraction or phase contrast image must be preserved through the imaging optics.

In the present case we model a 1 pC electron bunch photoemitted from a flat cathode in a 3.5 cell RF gun operating at 9.3 GHz with a peak cathode field of 170 MV/m and RF phase of 80$^o$ at emission with \textsc{parmela} \cite{young2003parmela}.  The electron bunch is generated by a UV laser pulse with 30~fs full width and a parabolic spatial distribution with RMS size of 30~$\mu$m in order to produce a 3-dimensional ellipsoid in blow-out mode  \cite{luiten2004realize, musumeci2008blowout}. The normalized emittance is
	\begin{equation}
	{{\varepsilon }_{xn}}=\frac{1}{m_ec}\sqrt{\left\langle {{x}^{2}} \right\rangle \left\langle p_{x}^{2} \right\rangle -{{\left\langle x{{p}_{x}} \right\rangle }^{2}}}
		\label{eq:emittance}
\end{equation}
where $x$ is the transverse coordinate, $p_x$ is the transverse momentum, $m_e$ is the electron mass and $c$ is the speed of light, with an initial value ${\varepsilon}_{xn} = 9$ nm-rad, which is an aggressive assumption but is supported by recent measurements \cite{li2015}. The gun exit energy is 4.5~MeV.  A short 20~cm linac then accelerates the beam to 7~MeV and removes its time-energy chirp.  A solenoid magnet surrounding the gun collimates the electron beam resulting in an RMS spot size at the crystal of $ \sigma_x = 101$~$\mu$m with an angular distribution $\sigma_{x'} =  7.7$~\urad~which is more than one order of magnitude smaller than the Bragg angle.  Simulation results show bunch length expansion to an RMS length of 100~fs with a peak current of 3.2~A. 

{Prior to analyzing the formation of a transverse electron-density modulation, we describe the underlying physics of electron diffraction occurring in the target.}
For single crystal Si with a lattice spacing of $a = 5.43~\angstrom$ we can see from Bragg's law, $\lambda =2{{d}_\textrm{hkl}}\sin {{\theta }}$ where $\lambda$ is the electron wavelength ($1.66\times10^{-3}~\angstrom$ at 7~MeV)and ${d}_\textrm{hkl}$ are the interplanar spacings, that for the lowest order diffracted beam ($\textrm{hkl}=(111)$) $2{{\theta }}=0.528$ mrad with respect to the incident electron beam. This small diffraction angle proves advantageous as it limits aberrations in the downstream electron optics. In the two-beam approximation the sample thickness, $t$, required for relativistic electrons and the (111) Bragg peak is given by the normalized amplitude ${{\left| {{\varphi }} \right|}^{2}}={{\sin }^{2}}\left( \frac{\pi t}{{{\xi }}} \right)$, where ${{\xi }} = 100$~nm is the extinction length given by ${{\xi }}={\pi \sqrt{1-\beta^2} {{V}_{c}}}/{\lambda {{F}}}$  \cite{fultz2012transmission,hirsch1966electron} with a structure factor of ${{F}}=22.6~\angstrom$ and ${{V}_{c}}={{\left( 0.543~\text{nm} \right)}^{3}}$.
\begin{figure*}[t]
  \centering
  \subfloat[]{\includegraphics[trim=0.25cm 0.25cm 0.25cm 0.25cm, clip=true, width=0.4\textwidth]{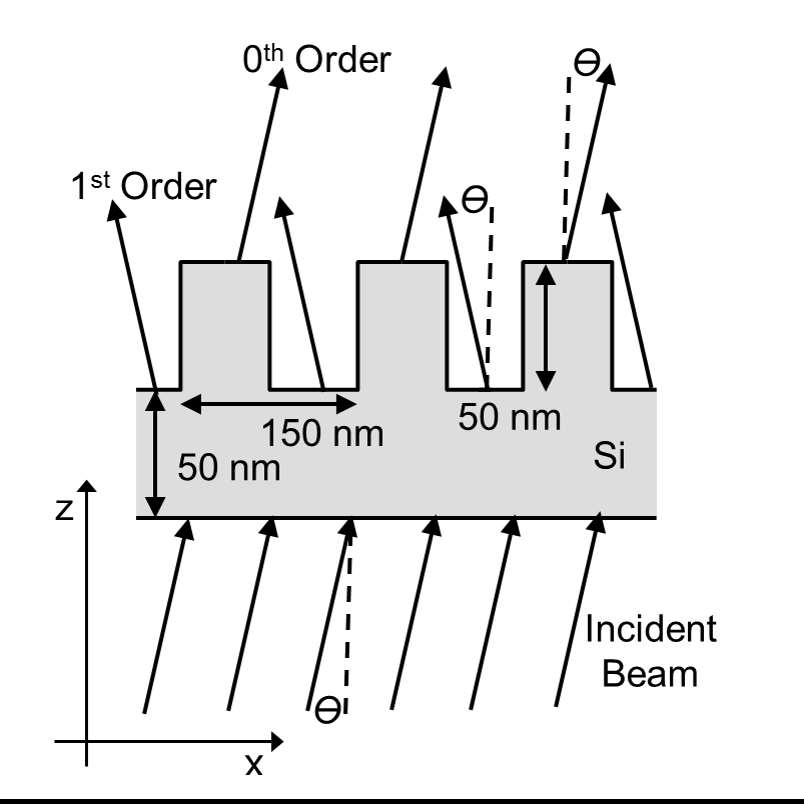}
	\label{fig:diffcontrast}}
	\subfloat[]{\includegraphics[width=0.45\textwidth]{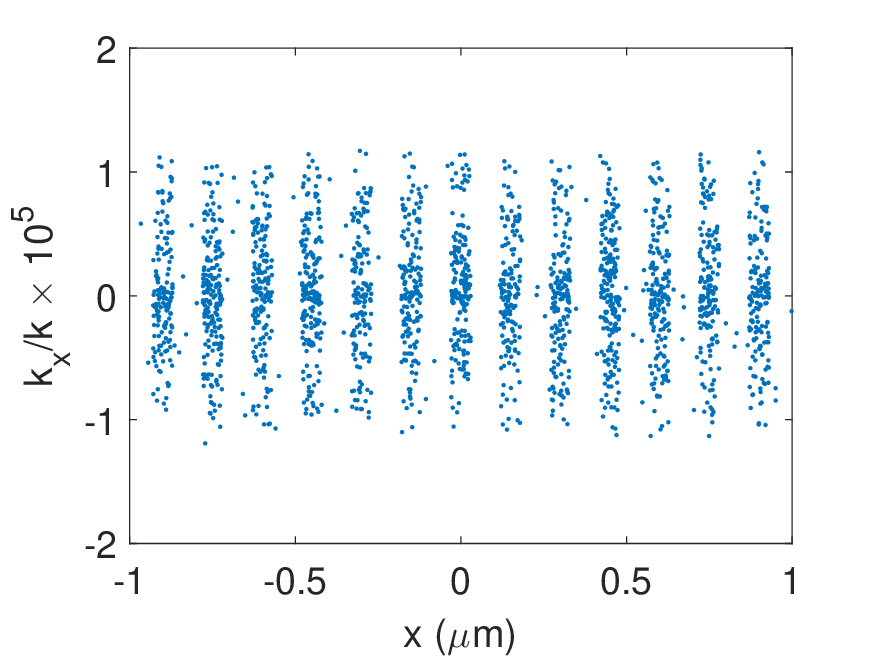}
	\label{fig:b0diffcontrast}} \\
	\caption {(a) The single crystal Si grating. (b) The forward scattered (0$^\textrm{th}$ order) electron beam {(with mean transverse momentum equal to the Bragg angle removed)}.  }
  \label{fig:grating}
\end{figure*}

\begin{figure}[t]
  \centering
  {\includegraphics[width=0.45\textwidth]{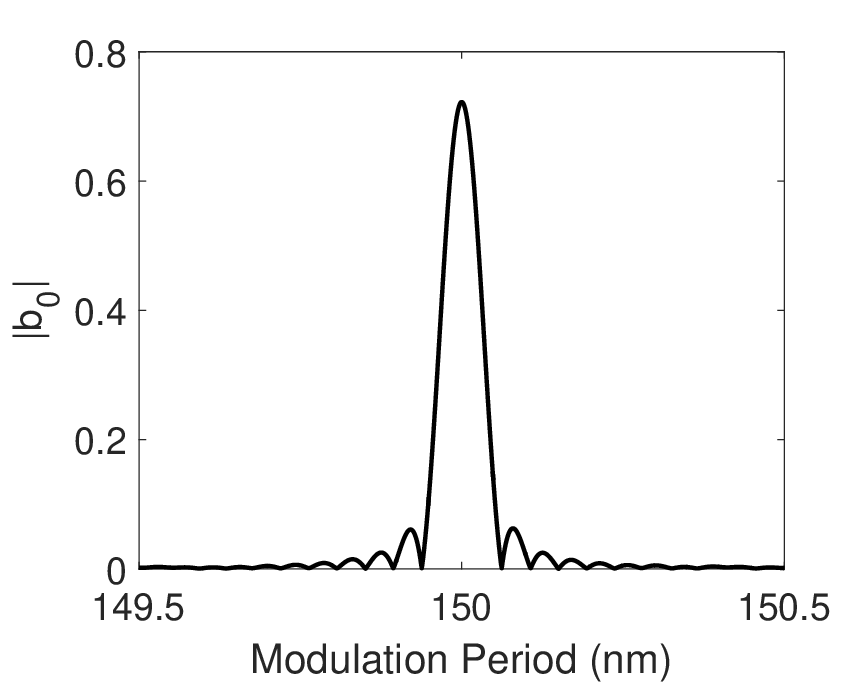}
  	\caption {The calculated $b_0$ with $\sim$150~nm spacing for the forward scattered (0$^\textrm{th}$ order) electron beam. }
  \label{fig:diffractioncontrast}}
\end{figure}

The scattering geometry is shown in \figref{fig:kspace}(a) with ${{\vec{k}}}+\vec{g}=\vec{k}'+\vec{s}$, where ${{k}}$ is the momentum vector for the incident electron, $k'$ is the momentum of a diffracted electron, $g=2\pi /d_\textrm{(111)}$ is the reciprocal lattice vector and ${s}$ is the deviation vector. Due to the finite emittance of the electron bunch, it is not possible for all the $k$ vectors of the incident particles to be properly aligned with the crystal plane, resulting in a a decreased probability of interacting with the crystal lattice for increasing ${s}$.

For clarity we will only consider the two beam case for a target consisting of a uniform medium. In practice, higher-order scattering events will be present requiring adjustments to sample dimensions and collection optics (bright-field vs dark-field). Targets consisting of layered materials can also be considered to produce the desired beam. 

 The intensity of the diffracted beam is
	\begin{equation}
	{{I}}=\frac{\sin {{\left(\pi t{{s}_\text{eff}} \right)}^{2}}}{\xi ^{2}s_\text{eff}^{2}}
	\label{eq:diffint}
	\end{equation}
where ${{s}_\text{eff}}=\sqrt{{{s}^{2}}+\xi^{-2}}$ and ${{I}_{0}}=1-{{I}}$ is the intensity of the forward beam\cite{howie1966diffraction}. \figref{fig:kspace}(b) depicts the electron bunch phase space incident on the target. The forward and diffracted beams for a uniform thickness single Si crystal are shown in \figref{fig:kspace}(c). These calculations are validated by electron diffraction experiments \cite{nanni2016measurements,malin2017a,zhang2017a} using electron bunches at MeV energies with comparable emittances to these simulation parameters and single crystal silicon membranes that have demonstrated the near total scattering of an electron bunch into a single Bragg peak.   

\begin{figure*}[ht]
  \centering
  \subfloat[]{\includegraphics[width=0.48\textwidth]{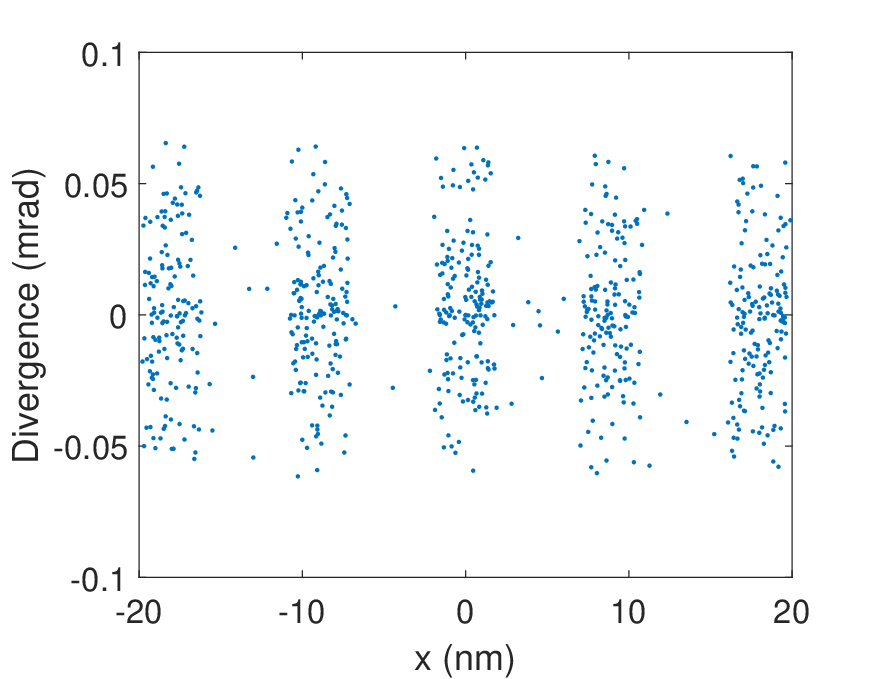}
  \label{fig:magmod}}
	\subfloat[]{\includegraphics[width=0.50\textwidth]{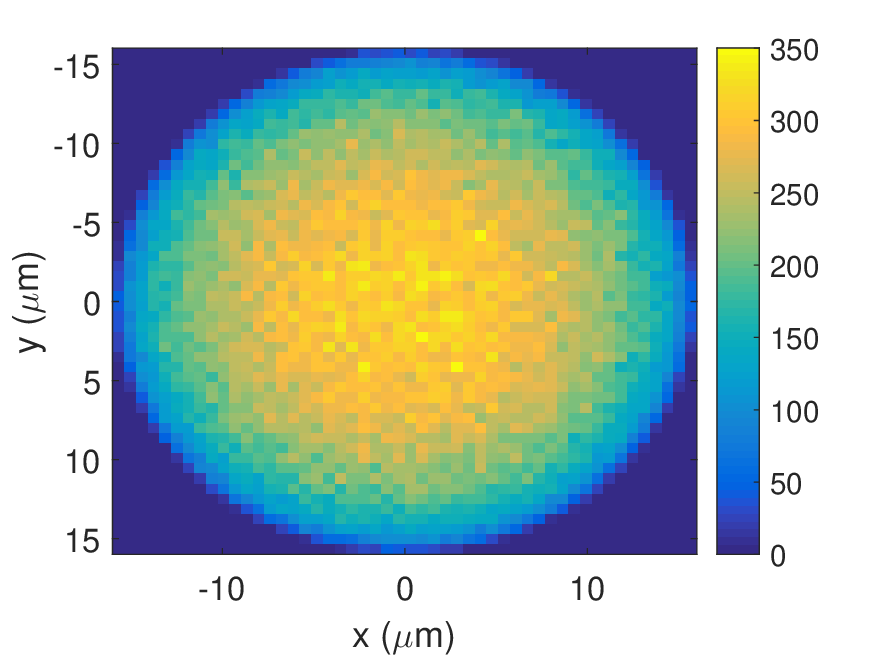}
	\label{fig:local_density}}
	\caption {(a) Individual electrons imaged after magnification and acceleration at the entrance of the EEX line. (b) Full transverse density profile of the electron bunch visualized as a two dimensional histogram of particle count in the simulation. }
  \label{fig:picmodel}
\end{figure*}

\begin{figure*}[ht]
  \centering
  \subfloat[]{\includegraphics[width=0.47\textwidth]{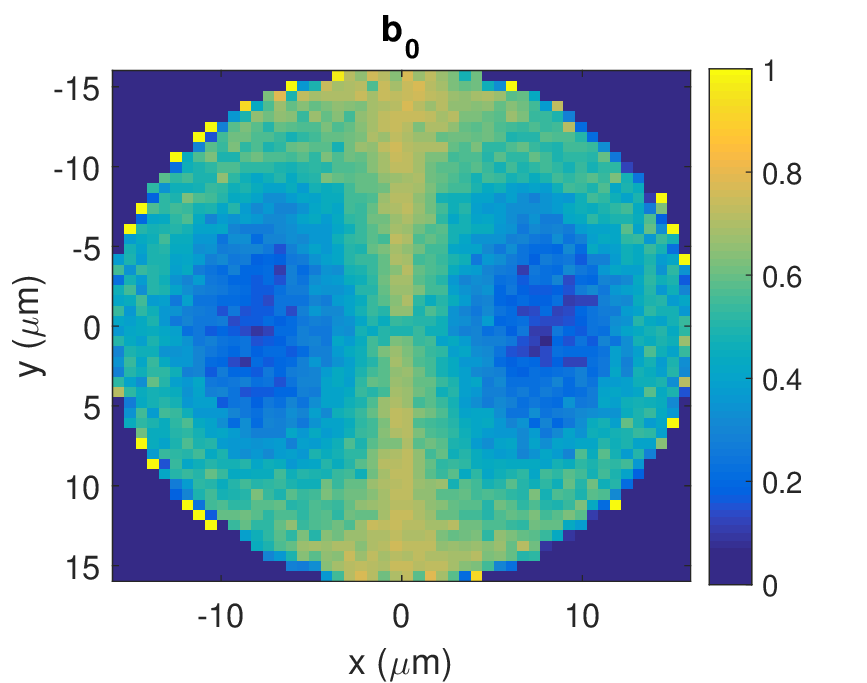}
  \label{fig:local_b0}}
	\subfloat[]{\includegraphics[width=0.50\textwidth]{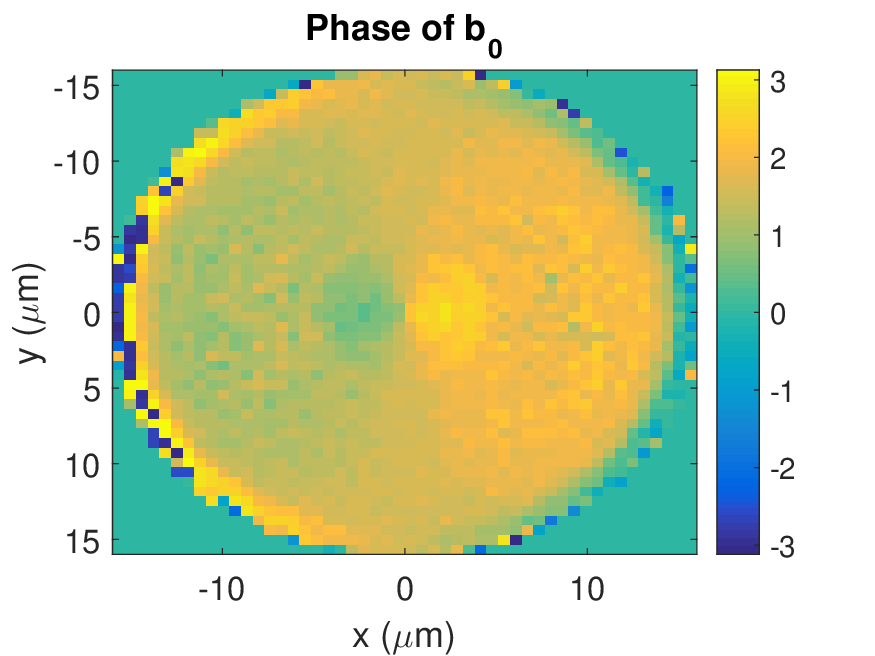}
	\label{fig:local_b0_phase}}
	\caption {(a) The bunching factor (b)  and the relative phase in radians as a function of the transverse position in the beam. }
  \label{fig:picmodel2}
\end{figure*}

\section{Diffraction Contrast Modulation}
\label{sec:difcontrast}

One approach to producing modulation in the electron bunch is to use a grating, \figref{fig:grating}(a). From \eqnref{eq:diffint} we can see that varying the thickness, $t$, in $x$  results in spatially alternating  intensities in the transverse dimension for ${{I}_{0}}$ and ${{I}}$, \textit{i.e.} diffraction contrast. Note that in this approach, the electrons are transmitted through the grating, and the grating is producing contrast due to the difference in diffracted intensity depending on the thickness of the grooves and ridges, but the diffraction itself is due to the Si atomic structure rather than the grating. The limitation on this modulation period is set by the minimum feature size that can be fabricated, however note that the image produced by the diffracted electron beam may be significantly demagnified to reach shorter periods. Subsequently, one of the two beams ($I_0$, $I$) could be blocked and the remaining beam would be sent through emittance exchange (EEX) optics \cite{cornacchia2002transverse,emma2006transverse,sun2010tunable,carlsten2011using,xiang2011emittance} to transfer the modulation from the transverse dimension to the longitudinal dimension. 

The collective quality of the modulated electron bunch is determined with the bunching factor, which is a useful tool to measure how well phased the modulation is at a particular wavelength. The bunching factor is defined as 
\begin{equation}
	{{b}_{0x}}=\frac{1}{{{N}_{e}}}\sum\limits_{p=1}^{{{N}_{e}}}{{{e}^{ik{{x}_{p}}}}}
\end{equation}	
where ${{N}_{e}}$ is the number of electrons, ${{x}_{p}}$ is the transverse position of the ${{p}^{th}}$ particle, $k={2\pi }/{{{\lambda }_{x}}}\;$, and ${{\lambda }_{x}}$ is the period of modulation.  For a random assortment of particles (no modulation) $b_{0x} = 1/\sqrt{N_e} \ll 1$. For the modeled case with a grating period of 150~nm, the forward scattered beam (\figref{fig:grating}(b)) contains 0.35~pC  and has a bunching factor $|b_{0x}| = 0.71$ (\figref{fig:diffractioncontrast}). The transverse modulation of either beam undergoes demagnification which can be varied and then is imaged into the longitudinal dimension via EEX.  We have recently shown \cite{nanni2015eex} that emittances with ratios as high as $10^4$ between longitudinal and transverse dimensions can be fully exchanged. If both beams are imaged without aberrations the modulation would disappear.

\begin{table*}[ht]
\caption{First Order Transfer Matrix \label{tab:rmatricies} }
\begin{center}
\begin{tabular}{l c c c }\hline\hline\
Element & Symbol& Analytical & Numerical \\
\hline
EEX Beamline & ${R}$  &$\left( \begin{matrix}
   0 & 0 & -7.27 & -0.0530  \\
   0 & 0 & -5.27 & -0.176  \\
   -0.176 & -0.043 & 0 & 0  \\
   -5.27 & -6.97 & 0 & 0  \\
\end{matrix} \right)$ & $\left( \begin{matrix}
   -1.13\times10^{-3} & -7.58\times10^{-4} & -6.92 & -0.034 \\
   -5.80\times10^{-3} &-1.05\times10^{-3} & -5.27 & -0.155  \\
   -0.174 & -0.029 & -3.72\times10^{-6} & -1.54\times10^{-5}  \\
   -5.33 & -6.63 & 1.72\times10^{-3} & -3.69\times10^{-4}  \\
\end{matrix} \right)$  \\

\hline \hline
\end{tabular}
\end{center}
\vspace{-3mm}
\end{table*}

\begin{figure*}[!ht]
  \centering
  \subfloat[]{\includegraphics[width=0.48\textwidth]{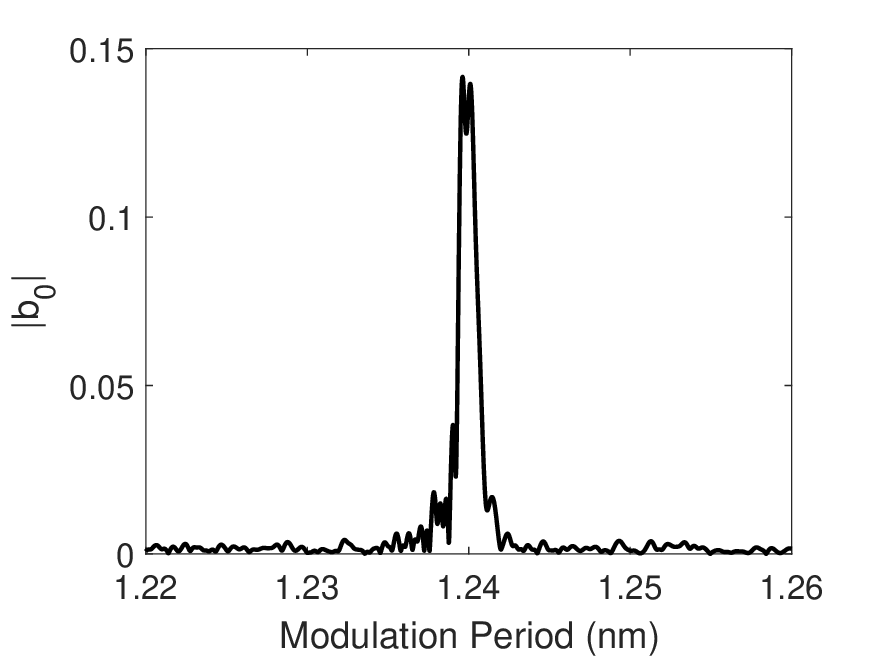}
  \label{fig:b0vslambda}}
	\subfloat[]{\includegraphics[width=0.48\textwidth]{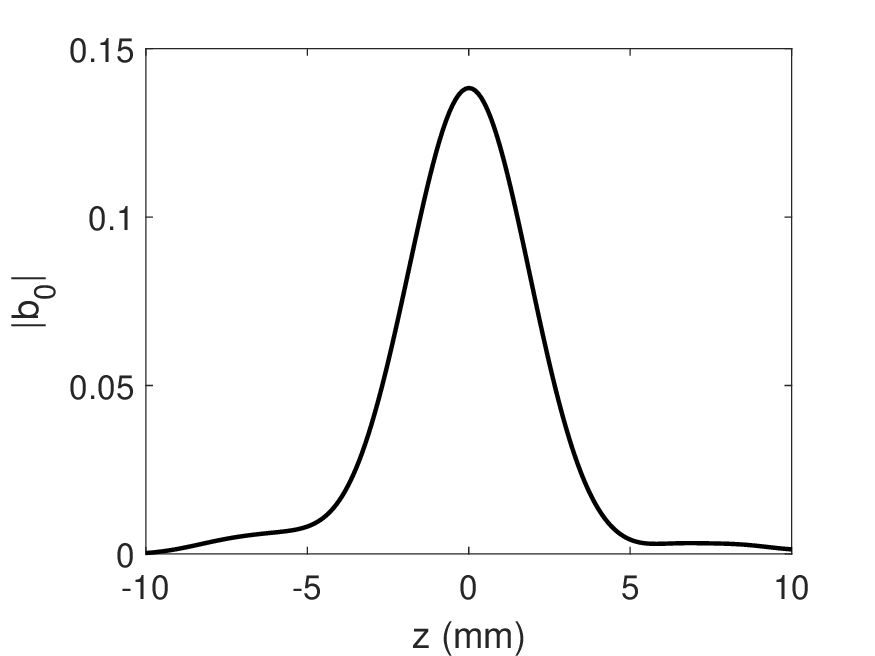}
	\label{fig:b0vsz}}
	\caption {(a) Individual electrons imaged after demagnification and acceleration at the entrance of the EEX line. (b) The bunching factor as a function of the transverse position in the beam.}
  \label{fig:picmodelEEX}
\end{figure*}

\section{Nanometer-Scale Longitudinal Modulation}

With an EEX beamline the emittance (phase-space) of an electron bunch in two dimensions is fully coupled and exchanged. For the beamline geometry we consider, the EEX line consists of two doglegs separated by a transverse deflecting cavity. With the modulation from the electron diffraction in the x-dimension, the EEX beamline is arranged such that EEX occurs between the x-z dimensions only; ideally the y-dimension is uncoupled. The behavior of the EEX beamline can be described by a transfer matrix $R$ applied to each electron's phase-space coordinates (shown without the y-dimension for simplicity)  
	\begin{equation}
			X^\text{T}=\left( \begin{matrix}  x &  x' & z & {\Delta p}/{p} \end{matrix} \right)
		\end{equation}
where all terms are defined with respect to a reference particle, $x$ is the transverse coordinates of the particle, $x'$ is the transverse angular divergence, $z=\beta c t$ is the longitudinal position, $p=\beta \gamma m_ec$ is the momentum, $c$ is the speed of light, $\beta=v/c$, $v$ is the electron velocity, $m_e$ is the electron mass and $\gamma=1/\sqrt{(1-\beta^2)}$ is the Lorentz factor. With the beamline arranged for complete emittance exchange the transfer matrix $R$ is 
	\begin{equation}
	R=\left( \begin{matrix}
   0 & 0 & -\frac{L}{\eta } & \eta -\frac{L\zeta }{\eta }  \\
   0 & 0 & -\frac{1}{\eta } & -\frac{\zeta }{\eta }  \\
   -\frac{\zeta }{\eta } & \eta -\frac{\zeta L}{\eta } & 0 & 0  \\
   -\frac{1}{\eta } & -\frac{L}{\eta } & 0 & 0  \\
\end{matrix} \right)
\end{equation}
where $\zeta$ is the longitudinal dispersion, $\eta$ is the horizontal dispersion and $L$ is the drift length \cite{nanni2015eex}. This transfer matrix will transfer the modulation from the transverse dimension into the longitudinal dimension.

The full accelerator setup is shown in \figref{fig:layout} including the interaction area where coherent x-rays would be produced via inverse Compton scattering (ICS) of a high power laser or THz field on the modulated electrons. A test case was analyzed for diffraction from a grating with 150~nm period and demagnification factor of 1/120 to produce a 1.24 nm modulation period which would produce coherent 1~keV x-rays. Diffraction contrast modulation with the Si structure and {dimensions} shown in \figref{fig:grating}(a) was used. After interacting with the Si grating the forward scattered beam is accelerated in a  short x-band linac from 7~MeV to 22.5~MeV. The accelerator is described in \cite{graves2014ics} and the arrangement of the EEX line is given in \cite{nanninanometer, nanni2015eex}. The imaging quadrupole lenses are set up as a telescope that demagnifies the beam by a factor of 1/14, and then the EEX line further demagnifies by a factor of 1/6.  These demagnification factors in combination with the modest demagnification due to acceleration result in a 1.24 nm modulation at the EEX output. \figref{fig:picmodel}(a) shows the imaged pattern of the Si grating after being accelerated and magnified with $b_{0x}=0.61$ for the entire electron bunch. The full transverse density profile of the electron bunch is shown in \figref{fig:picmodel}(b). The bunching factor calculated locally as a function of transverse coordinates is shown in \figref{fig:picmodel2}(a) along with the relative phase between the \figref{fig:picmodel2}(b) electron beam modulation showing a flat phase profile across the full bunch.

In practice emittance exchange suffers from some aberrations which can limit the ability of the EEX line to transfer the modulation into the longitudinal dimension. In \cite{nanni2015eex} an aberration corrected geometry for the EEX line was found to have a transfer matrix with aberration-free performance with emittances differing by four orders of magnitude between the beamlet produced by each grating period and the longitudinal bunch. The aberration corrected geometry required the addition of only three sextupoles and one octupole. The analytical and numerical performance from \textsc{parmela} simulations of the EEX line is given in \tblref{tab:rmatricies}. The transfer matrix contains some residual self coupling for both the transverse and longitudinal dimension. The electron bunch is primarily sensitive to residual self coupling in the longitudinal dimension, because the emittance of each beamlet produced by individual grating periods must be exchanged with the full emittance of the longitudinal bunch. As a result, the performance of the EEX line was optimized to minimize the residual self coupling in the longitudinal dimension as this dimension will contain the small final emittance. 

In \figref{fig:picmodelEEX}(a) the final bunching factor of the electron bunch after EEX is shown vs the modulation period. The bunching factor has decreased from a $b_{0x}$=0.61 at the entrance of the EEX line to $b_{0z}$=0.14 due to residual self-coupling of the longitudinal phase space. However, the line width of \figref{fig:picmodelEEX}(a) is 0.01~$\angstrom$ which corresponds to a coherent content equivalent to the transform limited bandwidth of the electron bunch after the grating. This indicates that the modulation in the electron bunch has been uniformly demagnified without impacting the coherence. The modulation as a function of position in the electron bunch is shown in distance is shown in \figref{fig:picmodelEEX}(b) with a strong coherent content throughout the longitudinal extent of the electron bunch.

\begin{figure*}[t]
  \centering
  \subfloat[]{\includegraphics[width=0.48\textwidth]{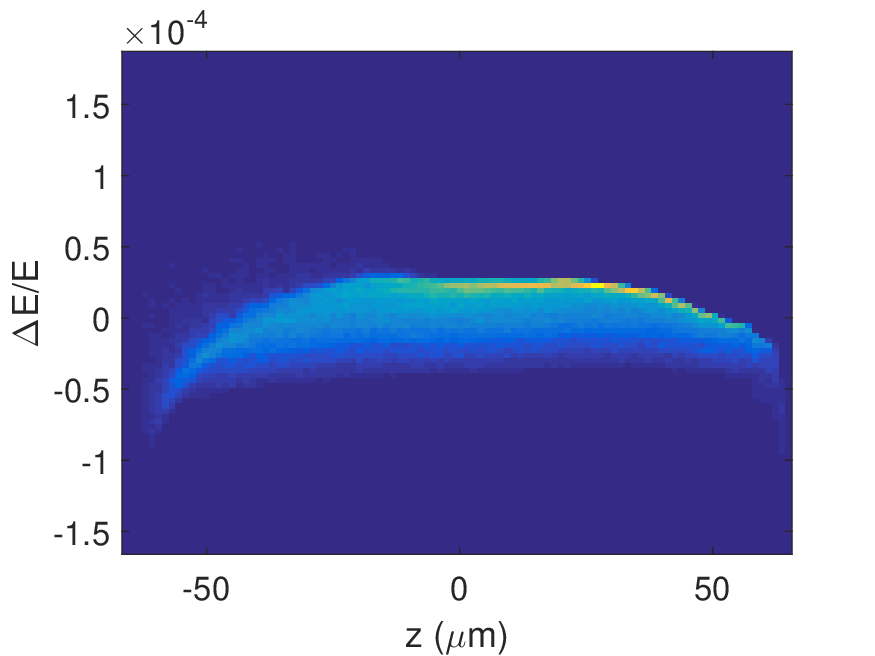}
  \label{fig:energyspreadbefore}}
	\subfloat[]{\includegraphics[width=0.48\textwidth]{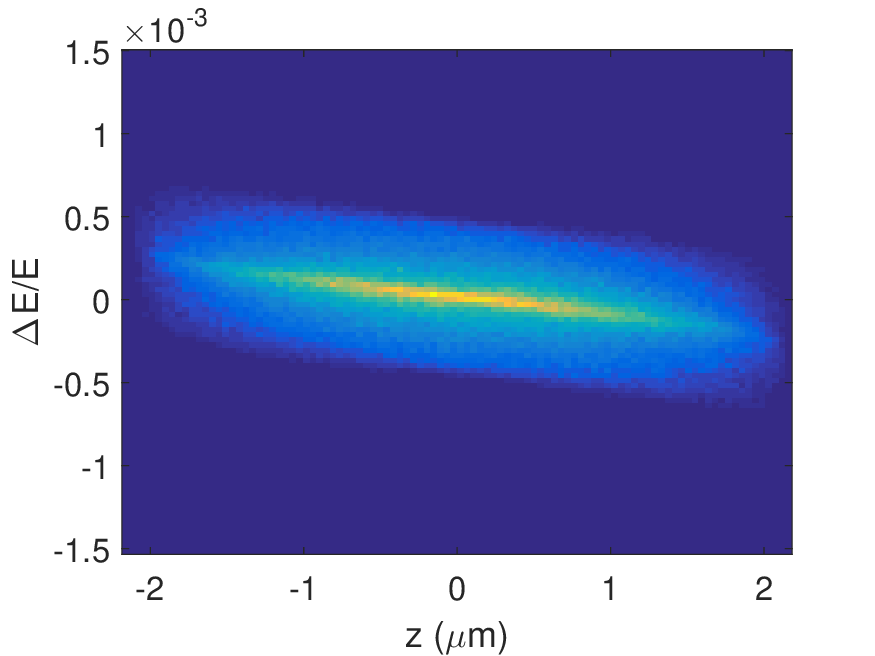}
	\label{fig:energyspreadafter}}
	\caption {The relative density of the electron bunch for the longitudinal phase space (a) before and (b) after emittance exchange.}
  \label{fig:energyspread}
\end{figure*}

\begin{figure}[t]
  \centering
  {\includegraphics[width=0.48\textwidth]{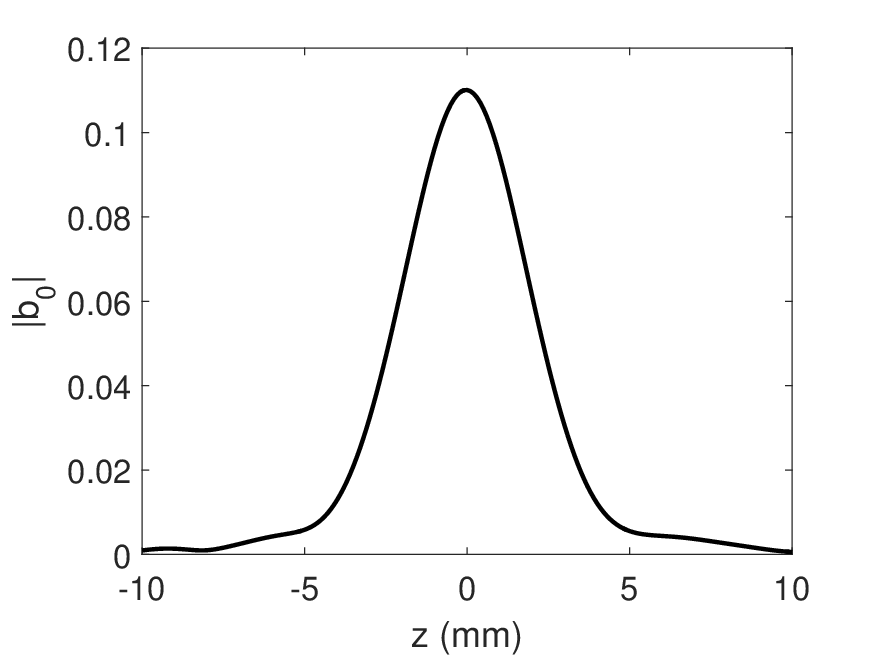}
  	\caption {The bunching factor for the electron bunch modulation at 1.24~nm after the EEX line. The depth of field is 4~mm, approximately ten times the gain length or four times the nominal laser pulse length.}
  \label{fig:density}}
\end{figure}

The impact of the emittance exchange line can also be observed in \figref{fig:energyspread} with the longitudinal phase space shown before and after the EEX line. The global phase space shows that there is no correlation before and after emittance exchange. The electron bunch length is compressed by the EEX line resulting in an increase in the peak current from a few amps to $\sim$50~A at the exit of the EEX line. This compression is beneficial because the electron bunch travels from the grating through the EEX line with low current limiting the impact of space charge. The depth of field for the modulated electron bunch is also critical in determining its ability to interact with the laser undulator. The global bunching factor is shown in \figref{fig:density} as the electron bunch propagates after the EEX line. Due to the low energy the depth of field for the modulation is only $\sim$5~mm, but this is sufficiently long to interact with the entire length of the laser undulator.

\section{X-Ray Performance}

We can estimate the 1 keV x-ray performance by treating the ICS interaction as an effective undulator field and applying the universal FEL scaling formulas of Xie \cite{xie2000fel}. The electron spot size at the crystal results in scattering from about 1800 grooves thus producing 1800 nanobunches after EEX for a total electron bunch length of 8~fs with peak current 47 A. The high-power IR laser pulse used for ICS is equivalent to a static undulator in this x-ray performance estimate.  This is physically correct if the laser field is uniform, a challenging issue for which solutions \cite{chang2013,seipt2015,lawler2013} have been proposed.  For an IR laser with 10 $\mu$m wavelength and strength parameter $a_0 = 0.4$, electron beam Twiss parameter $\beta_x = 2$~mm, normalized transverse emittance $\exn = 10$~nm, and energy spread $\Delta E/E = 2.5 \times 10^{-4}$, the effective Pierce parameter including the effects of emittance, diffraction, and energy spread \cite{Bonifacio1984} is $\rho = 3.2 \times 10^{-4}$, the exponential gain length including these effects is $L_g = 401$ $\mu$m, and the saturated power is 810 kW. Only 2-3 gain lengths are required to reach saturation because the electrons are bunched before interacting with the laser, allowing use of a few ps laser pulse.  Furthermore because the energy spread is an order of magnitude smaller than the Pierce parameter, the saturation power could be substantially exceeded by using a chirped laser pulse equivalent to a tapered undulator. 

\section{Phase Contrast Modulation}
\label{sec:phasecontrast}

In order to extend the electron bunch modulation into the hard x-ray regime (\textit{i.e.} sub-nm modulation) without significant amounts of demagnification, phase-contrast imaging of diffracted electrons can directly provide modulation on the order of the atomic structure spacing ($\sim5~\angstrom$). Phase-contrast imaging relies on the interference of the diffracted beam, ${{\phi }}$, with the forward scattered beam, ${{\phi }_{0}}$:
	\begin{equation}
	{{\varphi }_{0}}\left( r \right)={{\varphi }_{0}}\left( z \right){{e}^{i\vec{{{k}_{0}}}\cdot \vec{r}}}
	\end{equation}
	\begin{equation}
	{{\varphi }}\left( r \right)={{\varphi }}\left( z \right){{e}^{i\left( \vec{{{k}_{0}}}+\vec{g} \right)\cdot \vec{r}}}
	\end{equation}
	
\begin{figure*}[t]
  \centering
  \subfloat[]{\includegraphics[width=0.48\textwidth]{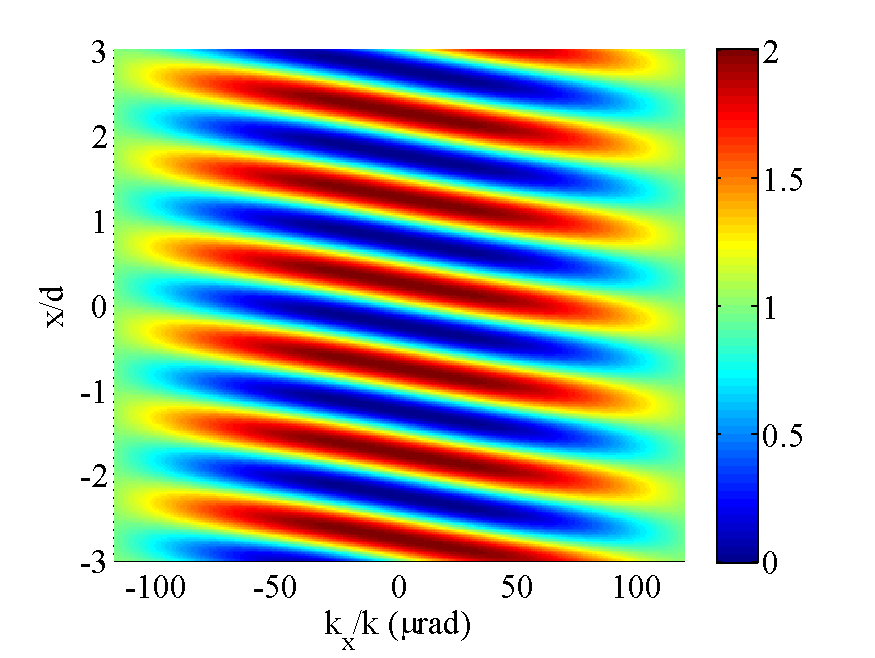}}
	\subfloat[]{\includegraphics[width=0.48\textwidth]{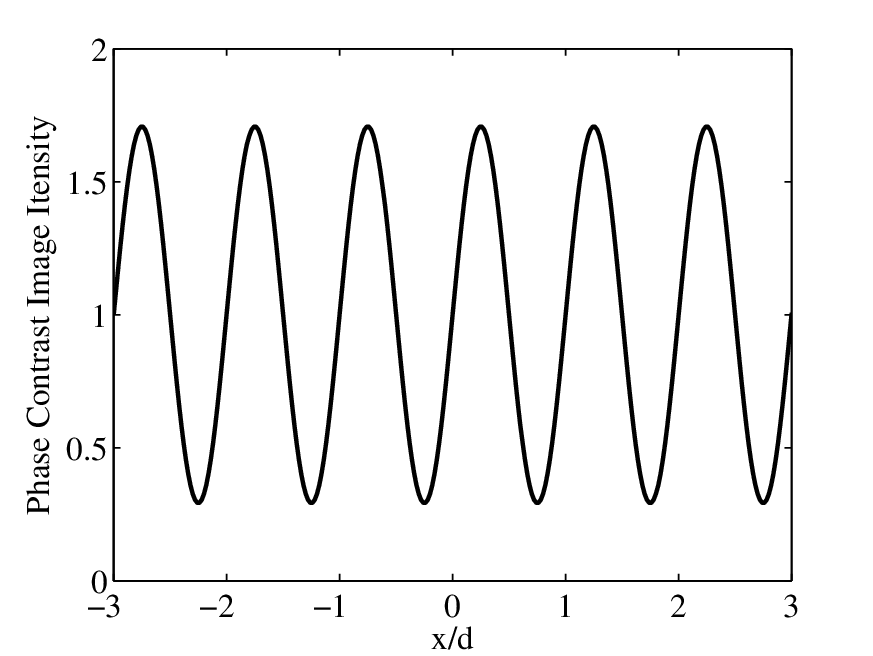}}
	\caption {(a) Phase-contrast modulation as a function of incident angle and transverse position. (b) Integrated phase-contrast modulation over the incident electron distribution ($\sigma_{x'} \approx  50$~\urad) at the exit of the Si crystal.}
  \label{fig:PhaseContrast}
\end{figure*}	
	
 The amplitude of these two wavefunctions is determined by the excitation of two Bloch waves $\left( {{\psi }_{1}}, {{\psi }_{2}} \right)$ at the entrance of the crystal and the relative phase of these two Bloch waves at the exit of the crystal. As the electron bunch arrives at the Si crystal, no modulation is present in the beam and its wavefunction is a plane wave with a flat phase front. Once the electron penetrates into the crystal it can no longer be described as a plane wave, because the Si atoms act as potential wells and apply a spatially varying phase advance. The Bloch waves $\left( {{\psi }_{1}}, {{\psi }_{2}} \right)$ are the new eigenstates for the electron, and the incident plane wave excites these two waves with equal amplitude for $\vec{s}=0$. Note that $\left( {{\psi }_{1}},{{\psi }_{2}} \right)$ propagate co-linearly, but they have unique wavevectors $( {{k}^{(1)}}, {{k}^{(2)}} )$ or $( \vec{k}+{{\gamma }^{(1)}}\hat{z}, \vec{k}+{{\gamma }^{(2)}}\hat{z} )$ . When the electrons exit the crystal we once again can describe them as plane waves with modulated phases. However, depending on the relative phase and amplitude of the two Bloch waves, two diffracted plane waves can be excited $\left( {{\phi }_{0}}, {{\phi }} \right)$. At the crystal exit the two waves are

\begin{eqnarray}\nonumber
{{\phi }_{0}}\left(z \right) & = & {{e}^{i{{s}}z/2}}\left[ \cos \left( \frac{{{s}_\text{eff}}z}{2} \right) \right. \\
&& \left.-i\cos \left( \cot {{\left( {{s}}{{\xi }} \right)}^{-1}} \right)\sin \left( \frac{{{s}_\text{eff}}z}{2} \right) \right]
\end{eqnarray}
	
	\begin{align}
	{{\phi }}\left(z \right) & = i{{e}^{i{{s}}z/2}}\sin \left( \cot {{\left( {{s}}{{\xi }} \right)}^{-1}} \right) \sin \left( \frac{{{s}_\text{eff}}z}{2} \right)
	\end{align}
noting that in $\widehat{x}$ both waves contain modulation in phase on the order of $g$. We can describe the total beam image as
	\begin{equation}
	{{I}_{tot}} =\left( {{\varphi }_{0}}+{{\varphi }} \right){{\left( {{\varphi }_{0}}+{{\varphi }} \right)}^{*}},
	\end{equation}
	
which becomes

\begin{eqnarray}\nonumber
{{I}_\text{tot}} & = & 1+2\sin \left( \cot {{\left( {{s}}{{\xi }} \right)}^{-1}} \right) \sin \left( \frac{{{s}_\text{eff}}z}{2}  \right) \\
&&  \times \left[ \sin \left( {{g}}x \right)\cos \left(\frac{{{s}_\text{eff}}z}{2}  \right)  \right. \nonumber \\
&& \left.+\cos \left( {{g}}x \right)\cos \left( \cot {{\left( {{s}}{{\xi }} \right)}^{-1}} \right) \right. \nonumber \\
&& \left. \times\sin \left( \frac{{{s}_\text{eff}}z}{2} \right)   \right].
\end{eqnarray}

We evaluate this expression for a sample depth of 75~nm or ${{3\xi }}/4$ which results in the optimal mix (50/50) of the forward and diffracted beam. \figref{fig:PhaseContrast}(a) is the intensity of the diffraction pattern in the transverse coordinate for varying incident angle and (b) is the accumulated population for our design case focused to a RMS spot size at the crystal of $ \sigma_x = 11.5$~$\mu$m with angular distribution $\sigma_{x'} \approx  50$~\urad~to reduce the number of modulation periods. Excellent phase contrast is observed at the exit of the Si crystal.  

\begin{figure*}[t]
  \centering
  \subfloat[]{\includegraphics[width=0.48\textwidth]{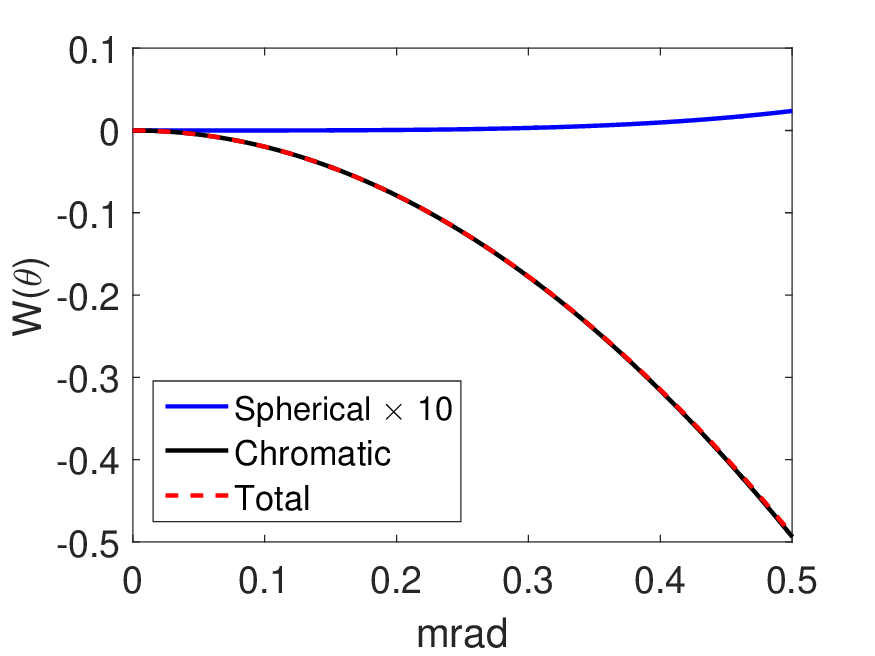}
  \label{fig:abb}}
	\subfloat[]{\includegraphics[width=0.48\textwidth]{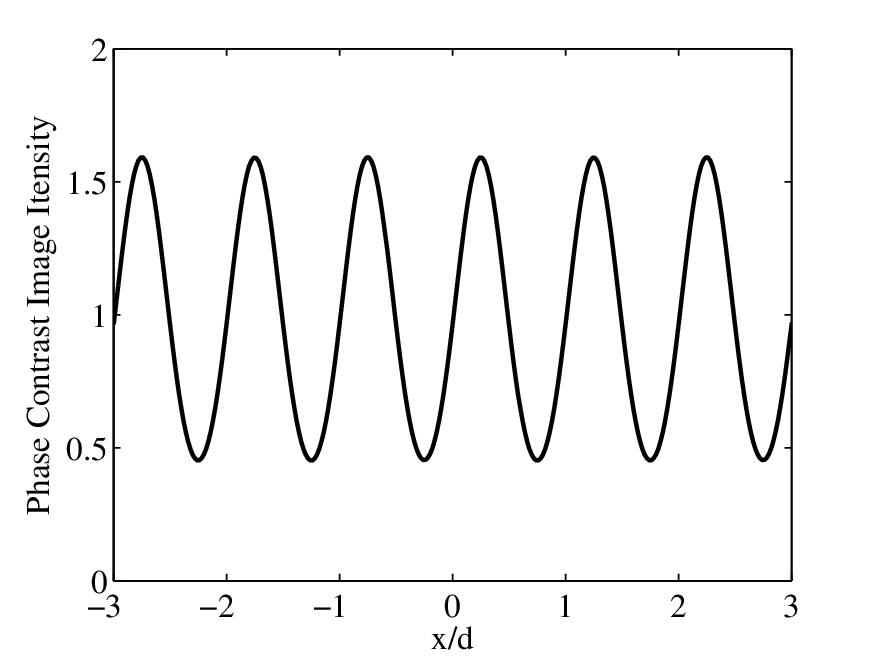}
	\label{fig:pc_mod}}
	\caption {(a) Added phase as a function of the angle of divergence for a focal length of 4~mm and a $\Delta E/E = 1.3 \times 10^{-5}$. (b) Electron population in phase-contrast image propagated through a lens with a focal length of 4~mm for the initial conditions in \figref{fig:PhaseContrast}(b).}
  \label{fig:pc_image}
\end{figure*}

\section{Aberrations}

Phase-contrast imaging has not yet been demonstrated with RF photo injectors, however the electron bunch produced by a state-of-the-art RF gun has sufficient beam quality as shown in the previous section. Imaging the transverse modulation at the interaction point will require careful analysis of aberrations from the optical elements in the setup. We estimate the effects of aberrations with analytical calculations that assume imaging with a standard objective lens. For phase contrast we must consider the wave-optical formulation of aberrations due to imaging. In the wave-optical formulation the effect of aberrations is given by a phase shift $W\left( \theta  \right)={2\pi \Delta s}/{\lambda }\;$ where $\Delta s$ is the change in optical path with respect to the ideal spherical wave front and $\theta$ is the scattered angle. The phase shift  can result from three effects: spherical aberrations, fluctuations in the thickness of the sample, or change in focal length due to energy. These effects combine to give a total phase shift (Eqn. 3.65 in \cite{reimer2008transmission}) of 
	\begin{equation}
	W\left( \theta  \right)=\frac{\pi }{2\lambda }\left( {{C}_{s}}{{\theta }^{4}}-2\left( \Delta f-\Delta a \right){{\theta }^{2}} \right)
	\label{eq:abphase}
	\end{equation}
where $\Delta f={f\Delta E}/{E}\;$ and $\Delta a$ is the variation of the longitudinal position of the sample (effectively due to tilt) and should be kept to on the order of $\Delta f$. It is sufficient to keep $\Delta a$ on the order of 40~nm (with an illumination spot of 11.5~$\mu$m this is a tilt of 4~mrad), which is a weaker tolerance than the required 0.1~mrad alignment for the crystal plane.

Observing modulation of the electron beam that is on the order of the lattice spacing  $a$ = 5.43~$\angstrom$ requires the ability to collect electrons from a transverse momentum space that covers ${{k}_{\bot }}={4\pi }/{a}\;$ or ${{{k}_{\bot }}}/{{{k}_{0}}=}\;0.5\text{ mrad}$ which includes a minimum of two diffraction peaks. To analyze the imaged beam we take the amplitude distribution at the output of the crystal $\phi \left( r,{{z}_{o}} \right)={{\phi }_{0}}\left( r,{{z}_{o}} \right)+{{\phi }}\left( r,{{z}_{o}} \right)$ and propagate it as spherical wave fronts to the image location \cite{reimer2008transmission} including aberrations from \eqnref{eq:abphase}. In the absence of aberrations, the objective lens will re-image the beam such that the relative accumulated phase at the image plane (${z}_\text{i}$) is zero, recreating the image
	\begin{equation}
	{{\phi }_{i}}\left( r,{{z}_\text{i}} \right)=\frac{1}{M}\iint{F(\mathrm{q}){{e}^{i2\pi \mathrm{q}\cdot \mathrm{r}}}}{{\text{d}}^{2}}\mathrm{q=}\frac{1}{M}\phi \left( r,{{z}_\text{o}} \right)
	\end{equation}
where $F\left( \mathrm{q} \right)=\int\limits_{S}{{{\phi }_{s}}\left( \mathrm{r} \right)}{{e}^{-}}^{i2\pi \mathrm{q}\cdot \mathrm{r}}{{\text{d}}^{2}}\mathrm{r}$ and $q={{{k}_{\bot }}}/{{{k}_{0}}}\lambda ;=\theta/\lambda $ is the transverse momentum. Aberrations given by the momentum space pupil function $H\left( \theta  \right)={{e}^{-iW\left( \theta  \right)}}M\left( \theta  \right)$ are included where the diaphragm opening $M\left( \theta  \right)$ is a step function describing the angle of rays which are collected for the image. The new imaging formulation is 
	\begin{equation}
	{{\phi }_{i}}\left( r,{{z}_\text{i}} \right)=\frac{1}{M}\iint{F(\mathrm{q}){{e}^{i2\pi \mathrm{q}\cdot \mathrm{r}}}}\text{H(}\mathrm{q}\text{)}{{\text{d}}^{2}}\mathrm{q},
	\end{equation}
which can also be described as a convolution of the source image with the pupil function for the objective lens $h\left( r \right)={{\text{F}}^{-1}}\left\{ H\left( \theta  \right) \right\}$, \textit{i.e.} the inverse Fourier transform of the momentum space pupil function. The amplitude distribution of the re-imaged electron beam is
	\begin{equation}
	{{\phi }_{i}}\left( r,{{z}_\text{image}} \right)=\frac{1}{M}\phi \left( r,{{z}_\text{o}} \right)\otimes h\left( r \right).
	\end{equation}
The aberrations and imaged electron beam are shown in \figref{fig:pc_image} assuming that the sample is placed a distance ${{S}_{1}}=2f$ from the objective lens.  \figref{fig:pc_image}(b) shows that a strong modulation is possible for the given beam and transport conditions. While this represents a first step in the analysis of a feasible arrangement for phase-contrast imaging, a significant amount of work remains in assessing the impact of aberrations introduced by the accelerating structures and the emittance exchange line. This will require operation at lower emittance or the addition of additional magnetic optics to reduce the impact of aberrations.
\\

\section{Conclusions}

In conclusion, we have presented detailed calculations showing how to produce a moderately relativistic electron beam that is coherently modulated at a nanometer scale in preparation for generating fully coherent x-rays.  Furthermore, we have presented a method to extend the technique to sub-nanometer modulation using phase contrast electron diffraction that would enable generation of coherent hard x-rays.  Using conventional ICS from a laser pulse, electron beams prepared in this manner can be a stable source of powerful fully coherent x-rays from a table-top source.  Such a source would have many impacts, enabling labs and groups around the world access at modest cost to the remarkable science produced by ultrashort pulse coherent x-rays.  The output pulse energy is modest due to the low charge employed, yet the temporal coherence and resulting spectral purity are likely to open new applications that are not achievable with SASE-based XFELs.  The coherent power significantly exceeds the startup noise in a SASE FEL and could be a useful seed source to transfer full coherence and improved stability to the x-ray beams produced large facilities.
  \\
\section{Acknowledgments}
\label{sec:acknowledgments}

 The authors gratefully acknowledge Lucas Malin, John Spence and Philippe Piot for many useful discussions. 
 
 This work was supported by NSF Grant No. DMR-1042342 and 1632780, DOE Grant No. DE-FG02-10ER46745 and  DE-AC02-76SF00515, and DARPA Grant No. N66001-11-1-4192.


\bibliography{PhDbib}
\bibliographystyle{apsrev}

\end{document}